\documentclass[reprint,showpacs,preprintnumbers,amsmath,amssymb,prc,floatfix]{revtex4-1}
\usepackage{color}
\usepackage{graphicx}
\usepackage{dcolumn}
\usepackage{bm}
\usepackage[colorlinks,citecolor=blue,linkcolor=red,anchorcolor=blue,filecolor=blue,urlcolor=blue]{hyperref}

\begin{document}

\title{Fusion cross-section for Ni-based reactions within the relativistic mean field formalism}
\author{M. Bhuyan$^{1}$}
\email{Email: bhuyan@ita.br}
\author{Raj Kumar$^2$}
\email{Email: rajkumar@thapar.edu}

\bigskip
\affiliation{$^1$Instituto Tecnol\'ogico de Aeron\'autica, 12.228-900 S\~ao Jos\'e dos 
Campos, S\~ao Paulo, Brazil}
\affiliation{$^2$School of Physics and Materials Science, Thapar Institute of Engineering 
and Technology, Patiala-147004, Punjab, India}
\date{\today}

\begin{abstract}
In this theoretical study, we establish an interrelationship between the nucleon-nucleon 
interaction potential and the nuclear fusion reaction cross-sections at low energies. The 
axially deformed self-consistent relativistic mean field with non-linear NL3$^*$ force is 
used to calculate the density distribution of the projectile and target nuclei for fusion. 
The Wong formula is used to estimate the fusion cross-section and barrier distribution 
from the nucleus-nucleus optical potential for Ni-based systems, which are known for fusion 
hindrance phenomena. The results of the application of the so obtained nucleus-nucleus 
optical potential for the fusion cross-section from the recently developed relativistic 
$NN-$interaction (R3Y) are compared with the well-known, phenomenological M3Y effective 
$NN$ potential. We found a relatively good results from R3Y interactions below the barrier 
energies as compare to the M3Y potential concerning the experimental data. We also observe 
the density dependence on the nuclear interaction potential in terms of nucleon-nucleon 
optical potentials.
\end{abstract}


\maketitle

\section{Introduction}
Efforts have been devoted to determining the nature of the nucleon-nucleon interaction 
since 1932, the discovery of the neutron by Chadwick as the heart of nucleus in nuclear 
physics \cite{brue53,alco72,sat79}. In the simplest expression, the nucleon-nucleon 
($NN$) interaction is considered as central and to have a typical square-well, Gaussian 
or Yukawa potential of various ranges and strengths, which can obtain the observed phase 
shifts in elastic-scattering processes \cite{brue53,reid68,cott73}. The traditional goal 
of nuclear physics is to understand properties of atomic nuclei regarding the bare 
interaction between a pair of nucleons. Though substantial progress has taken place to 
understand it in some theoretical and experimental attempts, remains an open problem at 
present. A large number of interactions have been constructed via studying $NN$ scattering, 
but there exist extensive modifications in the scattering behavior due to the presence of 
surrounding nucleons in a nucleus \cite{hama62,lass62,reid68,day81,stok94}. Further, the 
reconstruction of $NN-$ potential through particle exchanges is made possible by the 
development of quantum field theory \cite{yuka35,epel09,gros11,grac01}. An effective 
phenomenological interaction has an appropriate form to study the nuclear structure and 
dynamics, which typically depend on the local density of nucleus.

At low energy, one can assume that the interaction potential between a pair of nucleons 
is instantaneous and therefore the concept of a substantial theory of nuclear forces, 
applicable to nuclear structure calculations \cite{wein79,wein90,kapl98,kapl98a}. The 
analytical derivation of potential through particle exchange is important to understand 
the nuclear force as well as structural properties via the nucleus-nucleus optical 
potential for the study of many nuclear aspects such as nuclear radioactivity, nuclear 
scattering, nuclear fission and fusion process \cite{bhu12,bidhu14,bhu14a}. More 
fundamental approach to $NN-$interactions at low energies has been formulated by Refs 
\cite{wein79,kapl98a,van99,ulf09,ekst13} in terms of an effective theory for 
non-relativistic nucleons. It involves a few basic coupling constants, which have been 
determined from nucleon scattering data at low energies. Furthermore, the new effective 
$NN-$interaction entitled R3Y potential \cite{bhu12,bidhu14,bhu14a} analogous to the M3Y 
form \cite{sat79} can be derived from the relativistic mean field Lagrangian. This 
interaction depends on the relativistic force parameters, the coupling constant among 
the interacting mesons and their masses \cite{bhu12,bidhu14,bhu14a}. One can see various 
potentials in more details and use them in future studies for some general and up-to-date 
views on the subject, look, for instance in Refs. \cite{epel09,gros11,grac01,ulf09,van99}. 
Further, the nucleus-nucleus optical potential is quite important in the studies of elastic 
scattering of light and heavy-ion (HI) systems, in particular for the simple one-dimensional 
barrier penetration model (BPM) of fusion reaction, the barrier energy, the radius, and 
curvature via nuclear potential and Coulomb potential \cite{gold38,bohl82,stil89}. 
A microscopic description is required for calculating the interaction potential that 
incorporate the physical process, which can significantly influences the fusion process. 
The widely used methods to obtain the nuclear potential by integrating an $NN-$interaction 
over the matter distributions of the two colliding nuclei and the approach is called the 
double folding model \cite{gold38,bohl82,stil89}. It produces the nucleus-nucleus optical 
potential for further use in various studies including the radioactive decays 
\cite{bhu12,bidhu14,bhu14a,zhan07,khoa00,mukh02}. 

At low energy, the system can fuse either by penetrating the interaction barrier or it 
must have sufficient energy to overcome Coulomb barrier to get absorb. In the present 
study, we have considered the Ni-based reactions i.e $^{64}$Ni + $^{64}$Ni, $^{64}$Ni 
+ $^{124}$Sn, $^{64}$Ni + $^{132}$Sn, $^{58}$Ni + $^{58}$Ni, $^{58}$Ni + $^{124}$Sn, 
and $^{58}$Ni + $^{132}$Sn as their fusion excitation functions are available 
experimentally and also known for fusion hindrance \cite{jian04,lesk85,lesk86,jian15,lian07,lian08,beck81,wolf87,wolf87a,kohl11}. Below the Coulomb barrier, the nuclear structure 
effects dominate the resulting fusion dynamics, whereas the centrifugal potential 
suppresses the structure effects at above barrier energies. The estimation of fusion 
characteristics of heavy ions at extreme sub-barrier energies are of great interest for 
understanding the reaction mechanisms in astrophysics and the synthesis of the superheavy 
nuclei \cite{vand92,bala98,rowl91}. Hence, it is one of the great interest at present to 
see the performance of relativistic R3Y potential along with the microscopic relativistic 
mean field density to estimating the nuclear interaction potential for the study of fusion 
reaction at low energies. The present calculations are limited to the spherical coordinate 
system to generate the nucleus-nucleus interaction potential. One may consider the coupling 
between fusion and other degrees of freedom to generate a multidimensional potential barrier, 
which enhanced the fusion probabilities \cite{vand92,bala98,rowl91,vaz81,stok80,stok80a}. 
More detail studies of the multidimensional fusion barrier and their effect on the fusion 
dynamics can find from the Refs. \cite{vand92,bala98,rowl91,vaz81,stok80,stok80a,jian02,vaz81,jian03,jian06,jian07,hagi03,jian14,das98,jian04,lesk85,lesk86,jian15,lian07,lian08,beck81,wolf87,wolf87a,kohl11}. 

This paper is organized as follows: In Sec. II we discuss the theoretical model for the 
relativistic mean field approach along with double folding procedure to get microscopic 
nucleus-nucleus optical potential. The Wong formula is also discussed in this section 
to study the fusion characteristics. Sec. III is assigned to the discussion of the results 
obtained from our calculation and of the possible correlation among the $NN$ potential and 
fusion cross-section. Finally, a summary and a brief conclusion are given in Sec. IV.

\section{Relativistic mean-field formalism}
At present, the quantum chromodyanmics (QCD) is not conceivable to describe the complete 
picture of the hadronic matter due to its non-perturbative properties. Hence, one needs 
to apply the perspective of effective field theory (EFT) at low energy, known as quantum 
hadrodynamics (QHD) \cite{bogu77,sero86,ring86}. The mean field treatment of QHD has been 
used widely to describe the nuclear structure and infinite nuclear matter characteristics 
\cite{bogu77,sero86,ring86,bhu13,bhu14,bhu09,bhu11,bhu15,bhu18,bhu18a,lala99c}. In the 
relativistic mean field approach, the nucleus is considered as a composite system of 
nucleons interacting through exchange of mesons and photons 
\cite{sero86,rein89,ring96,vret05,meng06,paar07}. Here, most of the computational effort 
is devoted to solving the Dirac equation and to calculate various densities. We have used 
the microscopic self-consistent relativistic mean field (RMF) theory as a standard tool to 
investigate fusion study via Wong formula. The form of a typical relativistic Lagrangian 
density for a nucleon-meson many body system, 
\cite{bogu77,sero86,lala99c,bhu09,bhu11,bhu15,bhu18,bhu18a,rein89,ring96,vret05,meng06,paar07,niks11,zhao12,lala09,bend03}
\begin{eqnarray}
{\cal L}&=&\overline{\psi}\{i\gamma^{\mu}\partial_{\mu}-M\}\psi +{\frac12}\partial^{\mu}\sigma
\partial_{\mu}\sigma \nonumber \\ 
&& -{\frac12}m_{\sigma}^{2}\sigma^{2}-{\frac13}g_{2}\sigma^{3} -{\frac14}g_{3}\sigma^{4}
-g_{s}\overline{\psi}\psi\sigma \nonumber \\ 
&& -{\frac14}\Omega^{\mu\nu}\Omega_{\mu\nu}+{\frac12}m_{w}^{2}\omega^{\mu}\omega_{\mu}
-g_{w}\overline\psi\gamma^{\mu}\psi\omega_{\mu} \nonumber \\
&&-{\frac14}\vec{B}^{\mu\nu}.\vec{B}_{\mu\nu}+\frac{1}{2}m_{\rho}^2
\vec{\rho}^{\mu}.\vec{\rho}_{\mu} -g_{\rho}\overline{\psi}\gamma^{\mu}
\vec{\tau}\psi\cdot\vec{\rho}^{\mu}\nonumber \\
&&-{\frac14}F^{\mu\nu}F_{\mu\nu}-e\overline{\psi} \gamma^{\mu}
\frac{\left(1-\tau_{3}\right)}{2}\psi A_{\mu}. 
\label{lag}
\end{eqnarray}
The $\psi$ are the Dirac spinors for the nucleons. The iso-spin and the third component 
of the iso-spin are denoted by $\tau$ and $\tau_3$, respectively. Here $g_{\sigma}$, 
$g_{\omega}$, $g_{\rho}$ and $\frac{e^2}{4\pi}$ are the coupling constants for $\sigma-$, 
$\omega-$, $\rho-$ meson and photon, respectively. The constant $g_2$ and $g_3$ are for the 
self-interacting non-linear $\sigma-$meson field. The masses of the $\sigma-$, $\omega-$, 
$\rho-$ mesons and nucleons are $m_{\sigma}$, $m_{\omega}$, $m_{\rho}$, and $M$, respectively. 
The quantity $A_{\mu}$ stands for the electromagnetic field. The vector field tensors for 
the $\omega^{\mu}$, $\vec{\rho}_{\mu}$ and photon are given by,
\begin{eqnarray}
F^{\mu\nu} = \partial_{\mu} A_{\nu} - \partial_{\nu} A_{\mu}  \\
\Omega_{\mu\nu} = \partial_{\mu} \omega_{\nu} - \partial_{\nu} \omega_{\mu} 
\end{eqnarray}
and
\begin{eqnarray}
\vec{B}^{\mu\nu} = \partial_{\mu} \vec{\rho}_{\nu} - \partial_{\nu} \vec{\rho}_{\mu}, 
\end{eqnarray}
respectively. From the above Lagrangian density we obtain the field equations for the Dirac 
nucleons and the mesons (i.e., $\sigma$, $\omega$, and $\rho$, field) as,
\begin{eqnarray}
&&\Bigl(-i\alpha.\bigtriangledown+\beta(M+g_{\sigma}\sigma)+g_{\omega}\omega
+g_{\rho}{\tau}_3{\rho}_3 +g_{\delta}\delta{\tau}\Bigr){\psi} = {\epsilon}{\psi}, \nonumber \\
&& \left( -\bigtriangledown^{2}+m_{\sigma}^{2}\right) \sigma(r)=-g_{\sigma}{\rho}_s(r)
-g_2\sigma^2 (r) - g_3 \sigma^3 (r),  \nonumber \\
&& \left( -\bigtriangledown^{2}+m_{\omega}^{2}\right) V(r)=g_{\omega}{\rho}(r), \nonumber \\
&& \left( -\bigtriangledown^{2}+m_{\rho}^{2}\right) \rho(r)=g_{\rho}{\rho}_3(r). 
\label{eqm}
\end{eqnarray}
In the limit of one-meson exchange, for static baryonic medium, the solution of single 
nucleon-nucleon potential for scalar ($\sigma$), and vector ($\omega$, $\rho$) fields are 
given by,
\begin{eqnarray}
&&V_{\sigma}=-\frac{g_{\sigma}^{2}}{4{\pi}}\frac{e^{-m_{\sigma}r}}{r}
+\frac{g_{2}^{2}}{4{\pi}}\frac{e^{-2m_{\sigma}r}}{r^2}
+\frac{g_{3}^{2}}{4{\pi}}\frac{e^{-3m_{\sigma}r}}{r^3}  \nonumber\\
&&{\hbox{and}}\nonumber\\
&&V_{\omega}(r)=+\frac{g_{\omega}^{2}}{4{\pi}}\frac{e^{-m_{\omega}r}}{r},\quad
V_{\rho}(r)=+\frac{g_{\rho}^{2}}{4{\pi}}\frac{e^{-m_{\rho}r}}{r}.
\label{eq:9}
\end{eqnarray}
The total effective $NN-$interaction is obtained from the scalar and vector parts 
of the meson fields. The recently developed relativistic $NN-$interaction potential 
analogous to M3Y form \cite{sat79} entitled R3Y- potential. Here the R3Y potential 
is derived for the NL3$^*$ force, which is able to predict the nuclear matter as 
well as the properties of the finite nuclei at very high isospin asymmetries 
\cite{sch51,bhu12,bhu13,bhu14,bhu14a,bidhu14,vaut69,bhu15,bhu18}. The relativistic 
effective nucleon-nucleon interaction ($V^{R3Y}_{eff}$) for NL3$^*$ force along 
with the single-nucleon exchange effects as \cite{bhu14,bidhu14,bhu14a,sat79,bhu12},
\begin{eqnarray}
V^{R3Y}_{eff}(r)&=&\frac{g_{\omega}^{2}}{4{\pi}}\frac{e^{-m_{\omega}r}}{r}
+\frac{g_{\rho}^{2}}{4{\pi}}\frac{e^{-m_{\rho}r}}{r}
-\frac{g_{\sigma}^{2}}{4{\pi}}\frac{e^{-m_{\sigma}r}}{r} \nonumber \\
&& +\frac{g_{2}^{2}}{4{\pi}}\frac{e^{-2m_{\sigma}r}}{r^2}
+\frac{g_{3}^{2}}{4{\pi}}\frac{e^{-3m_{\sigma}r}}{r^3}
+J_{00}(E)\delta(s).
\label{r3y}
\end{eqnarray}
On the other hand, the M3Y effective interaction, obtained from a fit of the G-matrix 
elements based on Reid-Elliott soft-core $NN-$interaction \cite{sat79}, in an oscillator 
basis, is the sum of three Yukawa's (M3Y) with ranges 0.25 $fm$ for a medium-range attractive 
part, 0.4 $fm$ for a short-range repulsive part and 1.414 $fm$ to ensure a long-range tail of 
the one-pion exchange potential (OPEP). The widely used M3Y effective interaction 
($V^{M3Y}_{eff}(r)$) is given by
\begin{equation}
V^{M3Y}_{eff}(r)=7999\frac{e^{-4r}}{4r}-2134\frac{e^{-2.5r}}{2.5r},
\label{m3y}
\end{equation}
where the ranges are in $fm$ and the strength in MeV. Note that Eq. (\ref{m3y}) represents the 
spin- and isospin-independent parts of the central component of the effective $NN-$interaction, 
and that the OPEP contribution is absent here. Comparing Eqs. (\ref{r3y}) and/or (\ref{m3y}), 
we find similarity in the behavior of the $NN-$interaction and feel that Eq. (\ref{r3y}) can 
be used to obtain the nucleus-nucleus optical potential. One can find more details in the 
Refs. \cite{bhu12,bhu14,bidhu14}. The nuclear interaction potential, $V_n (R)$, between the 
projectile (p) and the target (t) nuclei, with the respective RMF (NL3$^*$) calculated nuclear 
densities $\rho_p$ and $\rho_t$, is
\begin{eqnarray}
V_{n}(\vec{R}) & = & \int\rho_{p}(\vec{r}_p)\rho_{t}(\vec{r}_t)V_{eff} 
\left( |\vec{r}_p-\vec{r}_t +\vec{R}| {\equiv}r \right) d^{3}r_pd^{3}r_t, \nonumber \\
\label{optp}
\end{eqnarray}
obtained by using the well known double folding procedure \cite{sat79} for the M3Y and the 
recently developed R3Y interaction potential, proposed in the Refs. \cite{bhu12,bhu14,bidhu14}, 
supplemented by zero-range pseudo-potential representing the single-nucleon exchange effects 
(EX). Adding Coulomb potential $V_C(R)$ (=$Z_pZ_te^2/R$) results in Nucleus-Nucleus interaction 
potential $V_T (R)$ [$=V_n(R)+V_C(R)$], used for calculating the fusion properties. As, we 
know the pairing play an important role in the nuclear bulk properties including the density 
distribution of open-shell nuclei, one has to consider the pairing correlation in their ground 
states \cite{karat10}. In case of nuclei not too far from the $\beta$-stability line, one can 
use the constant gap BCS pairing approach reasonably good and simple to take care of pairing 
correlation \cite{doba84}. The present analysis includes the intermediate mass nuclei around 
the $\beta-$stability, hence we have taken the relativistic mean field results with BCS 
treatment for pairing correlation \cite{mad81,moll88,bhu09,bhu15,bhu18,bhu18a}.

\begin{figure}
\begin{center}
\includegraphics[width=1.0\columnwidth]{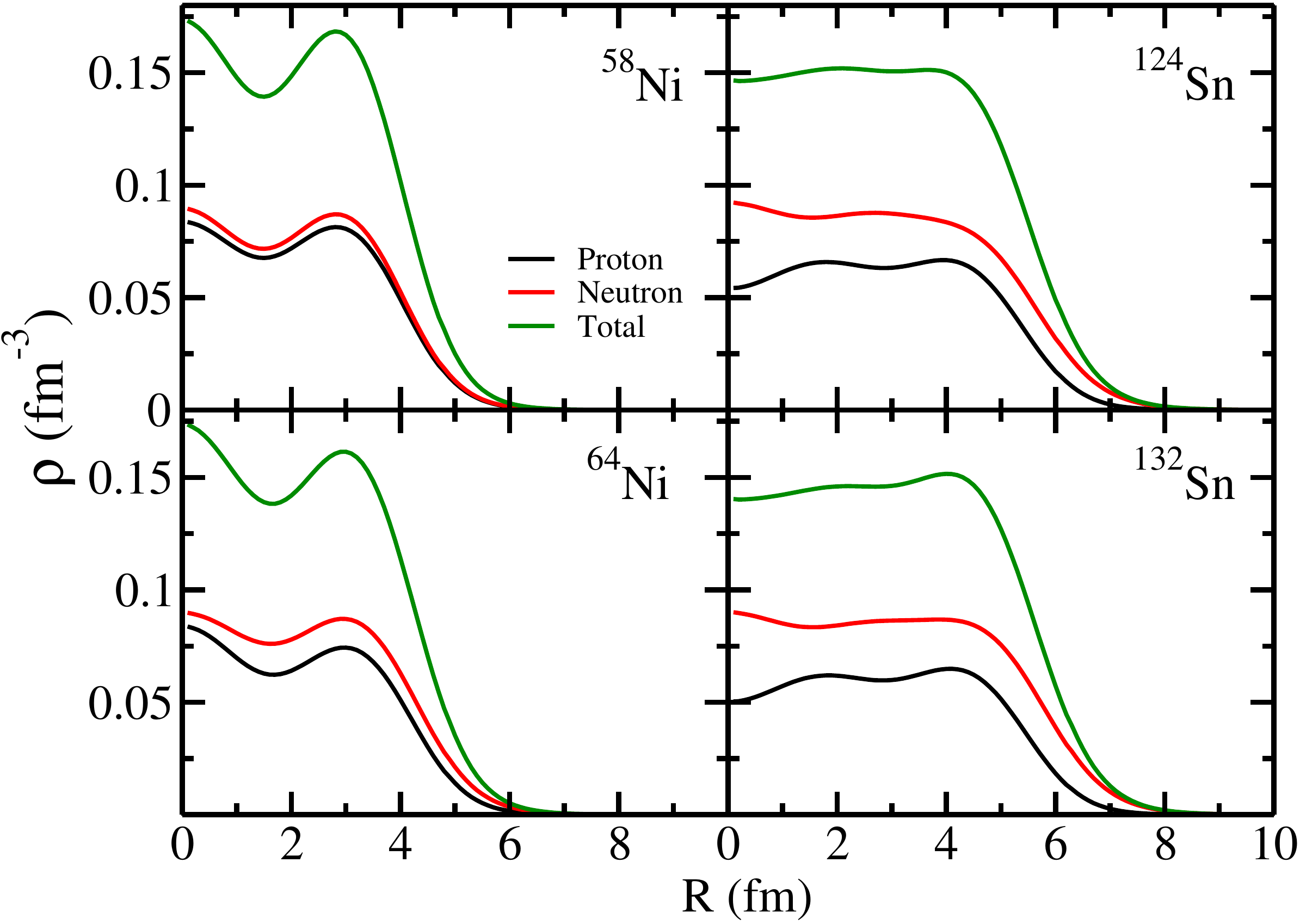}
\caption{\label{fig1} (Color online) The RMF (NL3$^*$) neutron, proton, and total radial 
density distribution for $^{58,64}$Ni, and $^{124,132}$Sn nuclei. See text for details.}
\end{center}
\end{figure}
\begin{figure}
\begin{center}
\includegraphics[height=10cm, width=9cm]{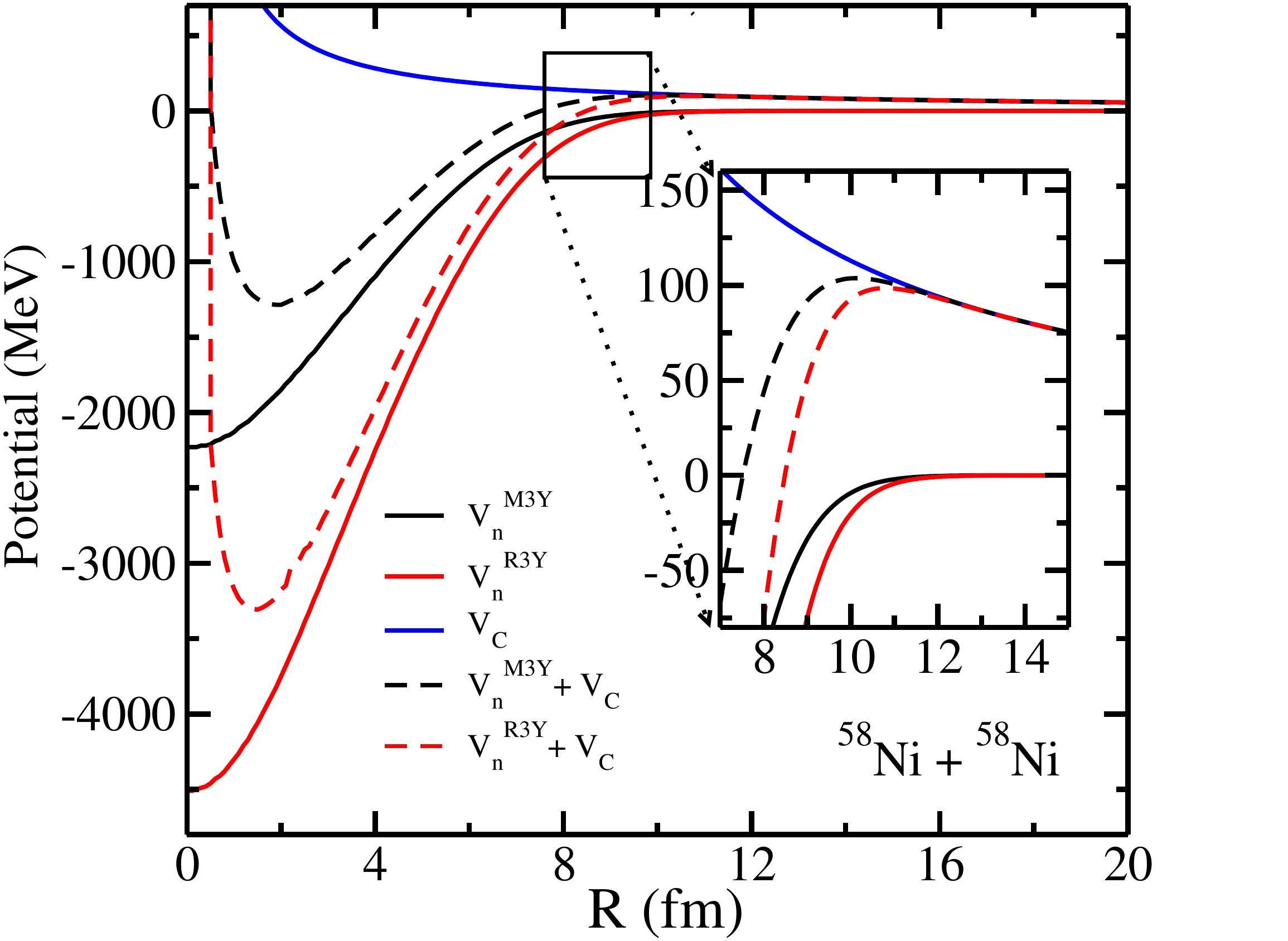}
\caption{\label{fig2} (Color online) The total nucleus-nucleus optical potential $V (R)$ 
and the individual contributions [the nuclear $V_n(R)(M3Y+EX)$ and $V_n(R)(R3Y+EX)$ for NL3 
parameter set, and the Coulomb $V_C(R)$ potential] as a function of radial separation $R$. 
The inset shows the barrier height and position in a magnified  scale.}
\end{center}
\end{figure}
\subsection{The Wong Formula}
In terms of $\ell$ partial waves, the fusion cross-section for two nuclei, colliding with
center-of-mass energy ($E_{c.m.}$), is given by \cite{wong73}
\begin{equation}
\sigma
(E_{c.m.})=\frac{\pi}{k^2}\sum_{\ell}(2\ell+1)P_{\ell}(E_{c.m.}),
\label{eq:23}
\end{equation}
with $k=\sqrt{\frac{2\mu E_{c.m.}}{\hbar^2}}$ and $\mu$ as the reduced mass. $P_{\ell}$ is 
the transmission coefficient for each $\ell$ which describes the penetration of barrier 
$V_T^{\ell}(R)$, given by 
\begin{equation}
V^{\ell}_{T}(R)=V_n(R,A_{i})+V_{C}(R,Z_{i})+\frac{\hbar^2\ell(\ell+1)}{2\mu
R^2}. \label{eq:1} 
\end{equation}
Using Hill-Wheeler \cite{hill53,thom59} approximation, the penetrability $P_{\ell}$, in terms 
of its barrier height $V_B^{\ell}(E_{c.m.})$ and curvature $\hbar\omega_{\ell}(E_{c.m.})$, is
\begin{equation}
P_{\ell}=\left[1+\exp\left(\frac{2\pi(V_B^{\ell}(E_{c.m.})-E_{c.m.})}
{\hbar\omega_{\ell}(E_{c.m.})}\right)\right]^{-1} \label{eq:24}
\end{equation}
with $\hbar\omega_{\ell}$ evaluated at the barrier position $R=R_B^{\ell}$ corresponding to 
barrier height $V_B^{\ell}$, given as
\begin{equation}
\hbar\omega_{\ell}(E_{c.m.})=\hbar\left[|d^2V_T^{\ell}(R)/dR^2|_{R=R_B^{\ell}}/\mu\right]^{1/2}
\label{eq:25}
\end{equation}
and, the $R_B^{\ell}$ obtained from the condition, $$|dV_T^{\ell}(R)/dR|_{R=R_B^{\ell}}=0.$$
Instead of solving Eq. (\ref{eq:23}) explicitly, which requires the complete $\ell$-dependent 
potentials $V_T^{\ell}(R)$, Wong \cite{wong73} carried out the $\ell$-summation in Eq. 
(\ref{eq:23}) {\it approximately} under specific conditions:\\ 
(i) $\hbar\omega_{\ell}\approx\hbar\omega_{0}$, and
(ii) $V_B^{\ell}\approx V_B^{0}+\frac{\hbar^2\ell(\ell+1)}{2\mu {R_B^{0}}^{2}}$,\\
which means to assume $R_B^{\ell}\approx R_B^0$ also. In other words, both $V_B^{\ell}$ and 
$\hbar\omega_{\ell}$ are obtained for $\ell$=0 case. Using these approximations, and replacing 
the $\ell$-summation in Eq. (\ref{eq:23}) by an integral, gives, on integration, the $\ell$=0 
barrier-based Wong formula \cite{wong73},
\begin{small}
\begin{equation}
\sigma(E_{c.m.})=\frac{{R_B^{0}}^{2}\hbar\omega_{0}}{2E_{c.m.}}.
\ln
\left[1+\exp\left(\frac{2\pi}{\hbar\omega_{0}}(E_{c.m.}-V_B^{0})\right)\right].
\label{eq:27}
\end{equation}
\end{small}
\vspace{-0.3cm}
\par\noindent
This is the simple formula used in the present work to calculate the fusion cross-section 
using the barrier characteristics such as V$_B^0$, R$_B^0$ and $\hbar\omega_{0}$ within the 
barrier penetration model for spherical nuclei. 
\begin{figure}
\begin{center}
\includegraphics[width=1.0\columnwidth]{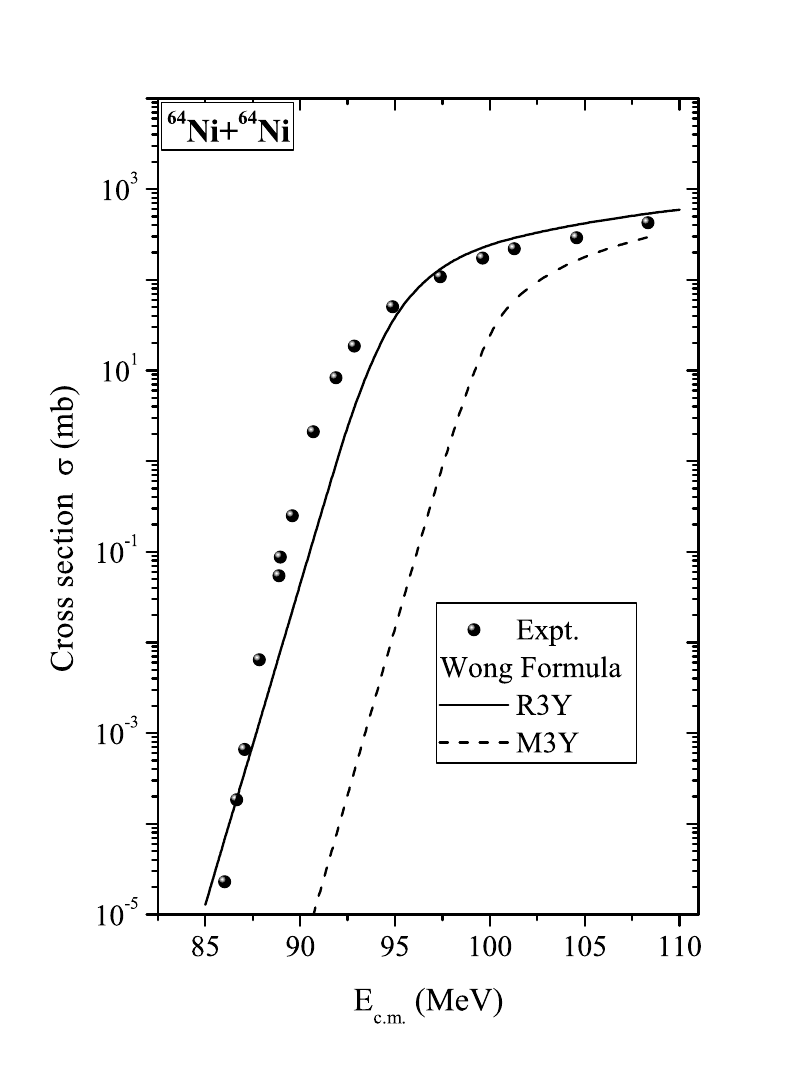}
\caption{\label{fig3} Fusion-evaporation cross-section as a function of center-of-mass 
$E_{c.m.}$, calculated by using the Wong formula for R3Y (solid line) and M3Y (dashed 
line) $NN-$interactions, and compared with experimental data for $^{64}$Ni+$^{64}$Ni 
\cite{jian04}. See the text for details.}
\end{center}
\end{figure}

\begin{figure}
\begin{center}
\includegraphics[width=1.0\columnwidth]{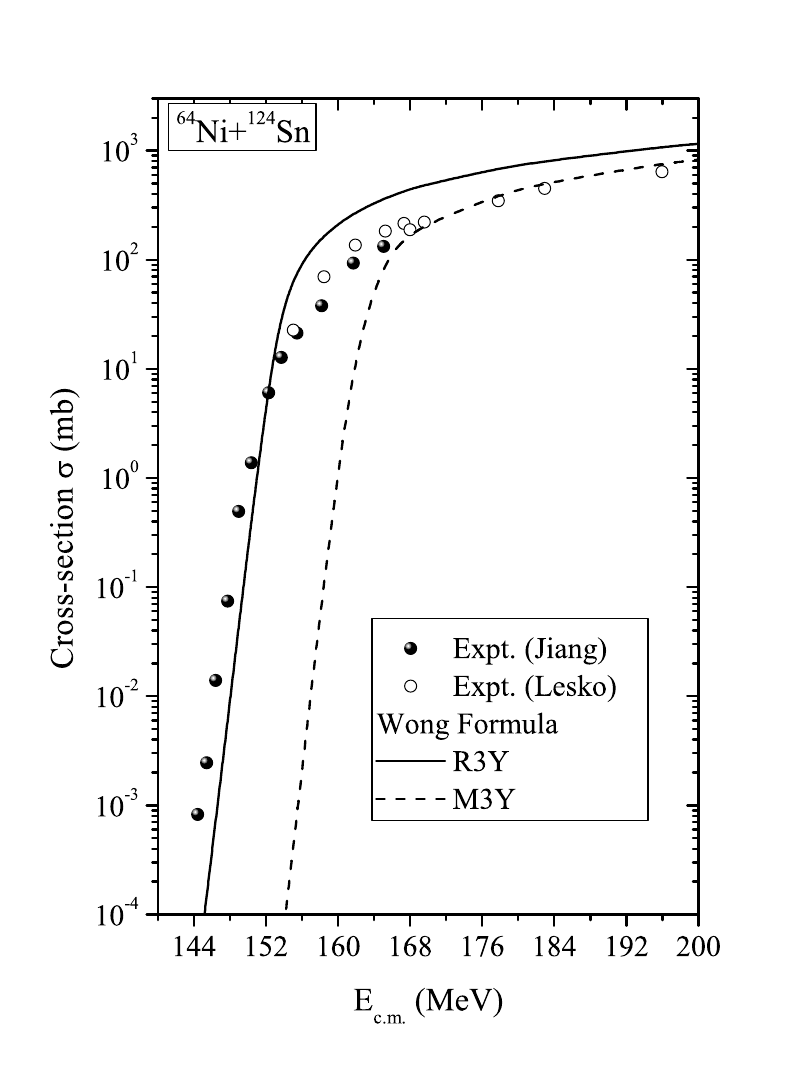}
\caption{\label{fig4} Same as for Fig. \ref{fig3}, but for the reaction of $^{64}$Ni 
+ $^{124}$Sn. The experimental data is taken from Refs. \cite{lesk85,lesk86,jian15}. 
See the text for details.}
\end{center}
\end{figure}
\begin{figure}
\begin{center}
\includegraphics[width=1.0\columnwidth]{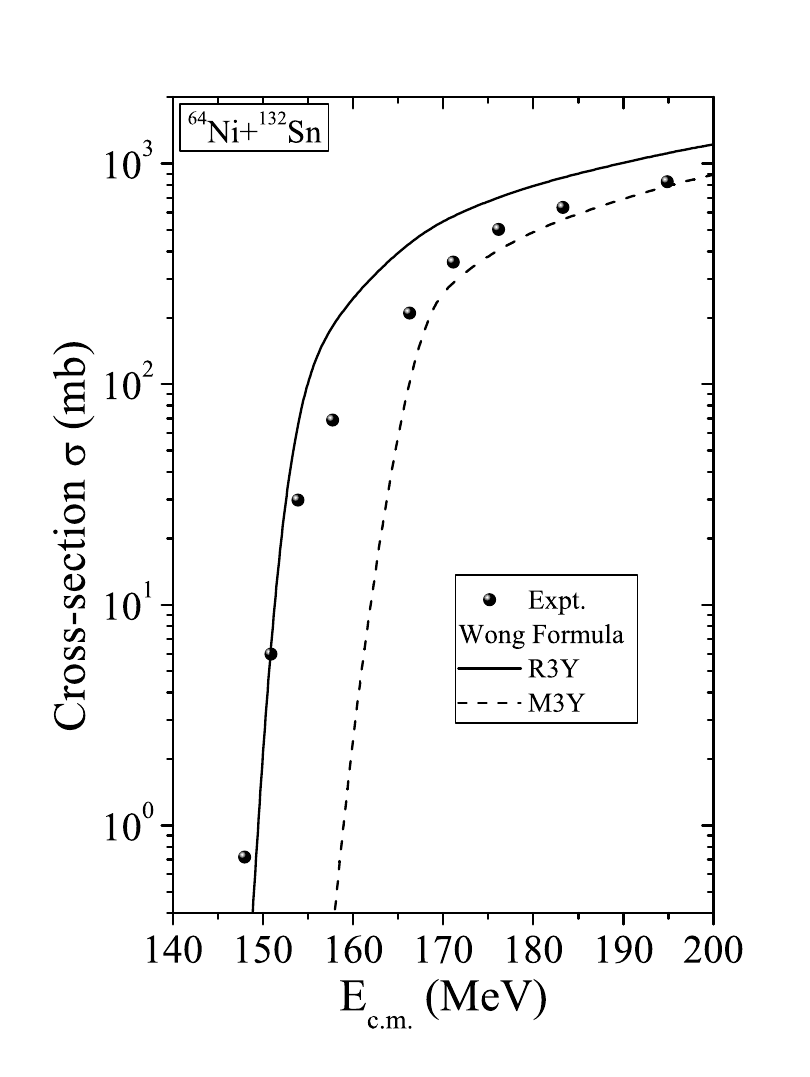}
\caption{\label{fig5} Same as for Fig. \ref{fig3}, but for the reaction of $^{64}$Ni 
+ $^{132}$Sn. The experimental data is taken from Refs. \cite{lian07,lian08}. See the 
text for details.}
\end{center}
\end{figure}
\begin{figure}
\begin{center}
\includegraphics[width=1.0\columnwidth]{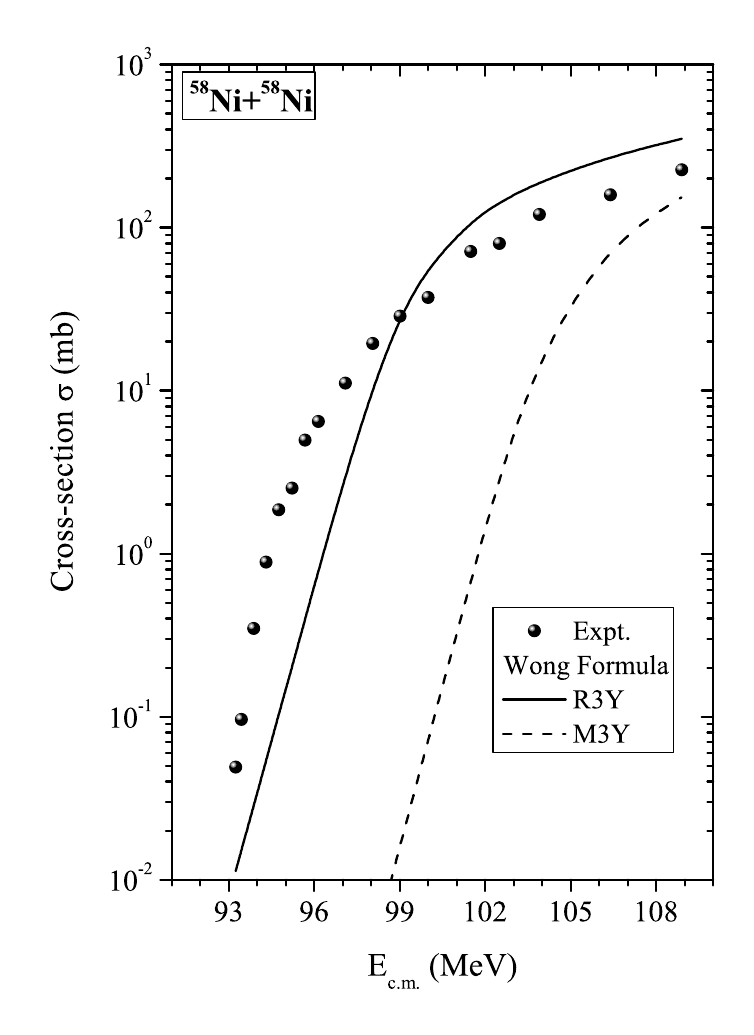}
\caption{\label{fig6} Same as for Fig. \ref{fig3}, but for the reaction of $^{58}$Ni 
+ $^{58}$Ni. The experimental data is taken from Refs. \cite{beck81}. See the text 
for details.}
\end{center}
\end{figure}
\begin{figure}
\begin{center}
\includegraphics[width=1.0\columnwidth]{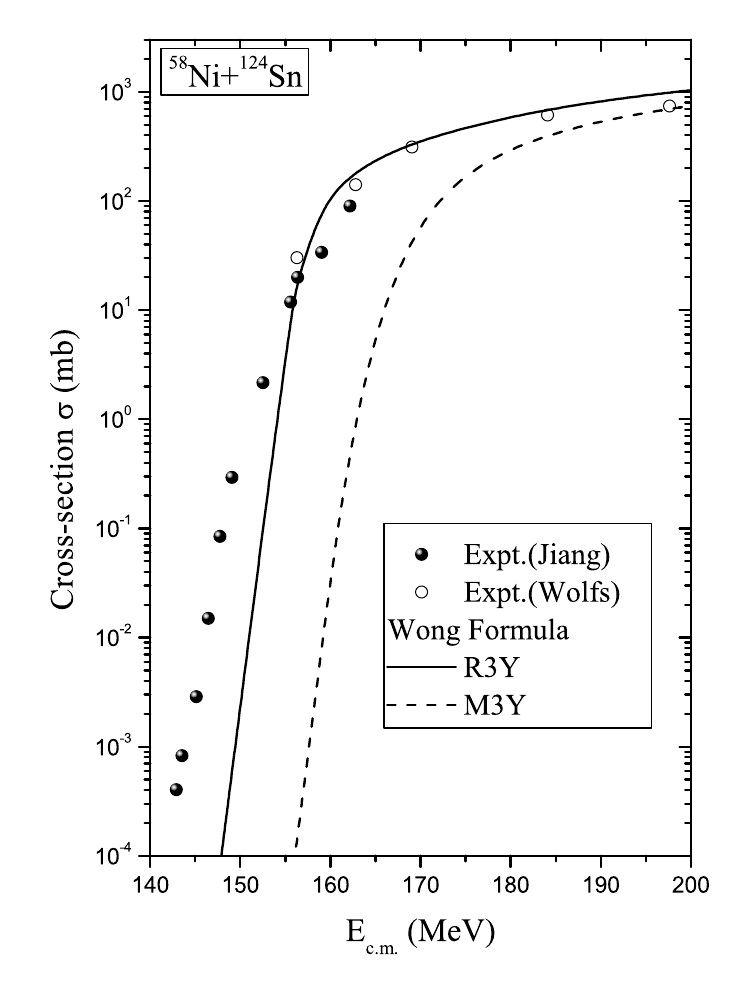}
\caption{\label{fig7} Same as for Fig. \ref{fig3}, but for the reaction of $^{58}$Ni 
+ $^{124}$Sn. The experimental data is taken from Refs. \cite{jian15,wolf87,wolf87a}. 
See the text for details.}
\end{center}
\end{figure}
\begin{figure}
\begin{center}
\includegraphics[width=1.0\columnwidth]{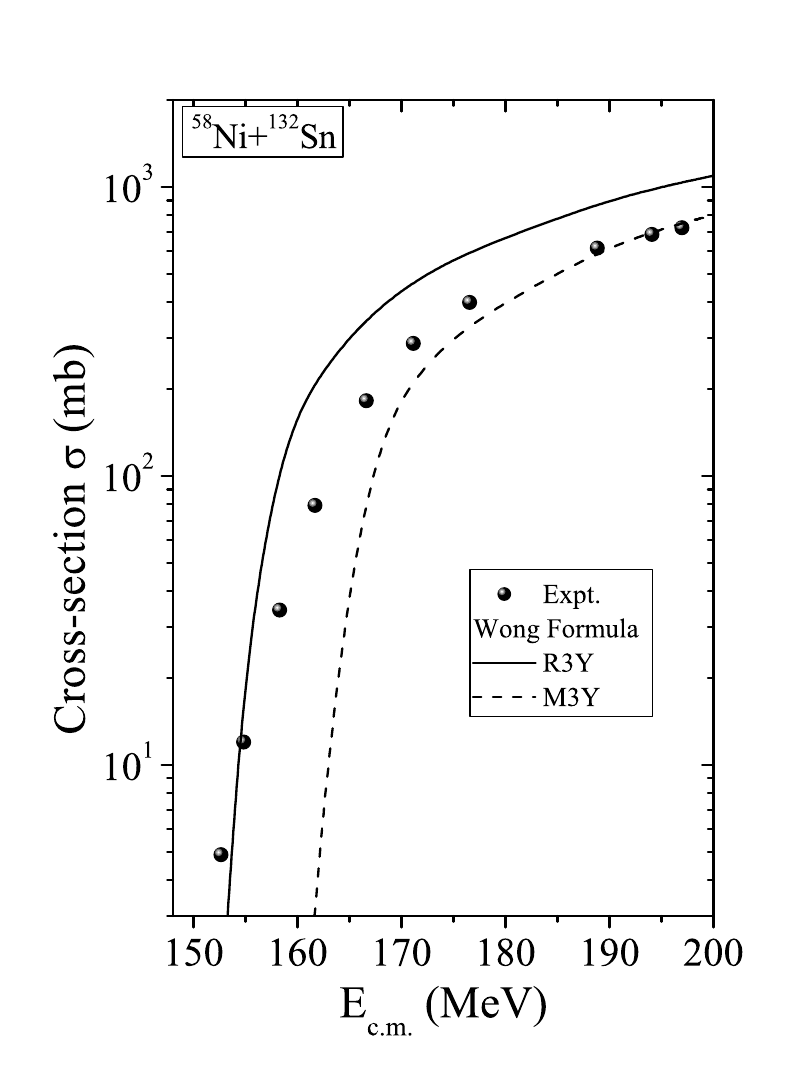}
\caption{\label{fig8} Same as for Fig. \ref{fig3}, but for the reaction of $^{58}$Ni 
+ $^{132}$Sn. The experimental data is taken from Refs. \cite{kohl11}.See the text for 
details.}
\end{center}
\end{figure}
\begin{figure}
\begin{center}
\includegraphics[height=10cm, width=9cm]{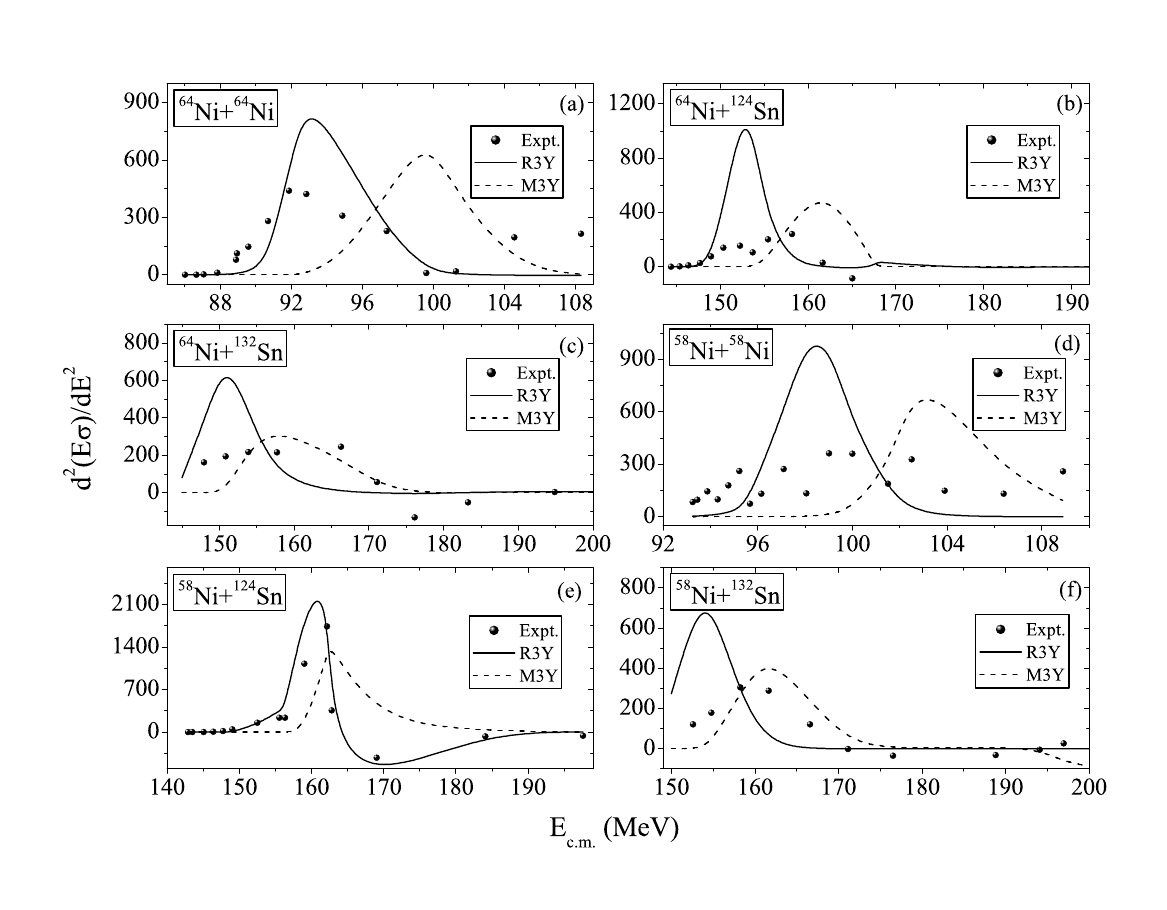}
\caption{\label{fig9} The barrier distributions for the reactions (a) $^{64}$Ni 
+ $^{64}$Ni, (b) $^{64}$Ni + $^{124}$Sn, (c) $^{64}$Ni + $^{132}$Sn, (d) $^{58}$Ni 
+ $^{58}$Ni, (e) $^{58}$Ni + $^{124}$Sn and (f) $^{58}$Ni + $^{132}$Sn as a function 
of center-of-mass energy. The experimental datas are taken from the Refs. 
\cite{jian04,lesk85,lesk86,jian15,lian07,lian08,beck81,wolf87,wolf87a,kohl11}. See 
the text for details.}
\end{center}
\end{figure}

\section{Calculation and Discussions}
The RMF calculations furnish principally the fusion hindrance reaction phenomena using the 
self-consistent relativistic mean field formalism via Wong formula. In the first step, we 
calculate the nuclear structure properties such as the binding energy, quadrupole moment 
$Q_{20}$, the total density distribution $\rho (r_{\perp},z)$ (i.e. the sum of the proton 
$\rho_p (r_{\perp},z)$ and neutron $\rho_n (r_{\perp},z)$ densities), the root-mean-square 
nuclear (neutron, proton and charge) radii and the single particle energy level for nucleons. 
Instead of concentrating on nuclear structure output profile for the NL3$^*$ force parameter, 
we use the monopole component of the densities for the target (t) and projectile (p) as the 
input for estimation of the optical nucleus-nucleus interaction potential using Eq. 
(\ref{optp}). The expression for the spin-independent proton and neutron mean-field densities 
from RMFT as, 
\begin{eqnarray}
\rho ({\bf R}) = \rho (r_{\perp}, z).  
\label{rhoR}
\end{eqnarray}
Here $r_{\perp}$ and $z$ are the cylindrical coordinates of the radial vector ${\bf R}$. The 
single particle densities are
\begin{eqnarray}
\rho_i ({\bf R}) = \rho_i (r_{\perp}, z) = \vert \phi_i^+ (r_{\perp}, z) \vert^2 + 
\vert \phi_i^- (r_{\perp}, z) \vert^2, 
\label{spd}
\end{eqnarray}
where, $\phi_i^{\pm}$ is the wave function, expanded into the eigen functions of an 
axially symmetric deformed harmonic oscillator potential in cylindrical co-ordinates. 
The normalization of the densities is given by,
\begin{eqnarray}
\int \rho ({\bf R}) d{\bf R} = X,
\label{norm}
\end{eqnarray}
where X = N, Z for neutron and proton number, respectively. Further, the multipole 
decomposition of the density can be written in terms of even values of the multipole 
index $\lambda$ as,
\begin{eqnarray}
\rho (r_{\perp}, z) = \sum_{\lambda} \rho_i ({\bf R}) P_{\lambda} (Cos\theta).
\label{mdd}
\end{eqnarray}
The monopole component of the density distribution of the expansion in the Eq. (\ref{mdd}) 
are used for calculating the nucleus-nucleus optical potential. In Fig. \ref{fig1}, we have 
plotted the neutron (black solid line), proton (red solid line) and total density (green 
solid line) distribution for $^{58,64}$Ni and $^{124,132}$Sn obtained from the NL3$^*$ 
force parameter as a function of radius. From the figure, one can find that the central 
density a bit smaller in magnitude and enhanced a little towards the surface region in case 
of $^{124,132}$Sn than that of $^{58,64}$Ni, which accepted to be the common feature in the 
heavy nucleus, play significant roles in the scattering studies \cite{bhu10}. 

\subsection{Nucleus-Nucleus Optical Potential}
The nuclear interaction potential V$_n$ (R) between the projectile (p) and target (t) 
nuclei is calculated using the well known double folding procedure in Eq. (\ref{optp}) 
\cite{sat79,bhu12} from respective RMF matter densities $\rho_p$ and $\rho_t$ for M3Y 
and recently developed relativistic R3Y $NN$ potential. The R3Y interaction is estimated 
for NL3$^*$ force parameter for present analysis \cite{bhu15,bhu18,bidhu14}, in which an 
effective Lagrangian is taken to describe the nucleon’s interaction through the effective 
mesons and electromagnetic fields. It is worth mentioning that the applicability of our 
newly introduced R3Y interaction potentials are used for the radioactivity studies of some 
highly unstable proton and/or neutron rich nuclei using preformed cluster decay model 
(PCM) of Gupta and co-workers \cite{bhu12,bhu14a,bidhu14}. The conservation of the angular 
momentum in the present analysis only limit to the ground state for estimate fusion 
characteristics of the constituent nuclei. To co-relate the theoretical calculation with 
the experimental fusion data, one need to adjust the spectroscopic factor  by including 
the particle vibration coupling. Nevertheless, without this particle vibration coupling 
our present formalism simply with the nonlinear $\sigma-$ mesons in Eq. (\ref{eq:9}), 
which is able to produce reasonable agreement with the experimental data \cite{bidhu14}. 
Furthermore, these non-linear terms in the $\sigma-$ field plays an important role in the 
nuclear matter study and the detailed nuclear structure inherited by the density while 
calculating the proton and cluster decay properties (mostly a surface phenomenon) 
\cite{bhu18,bhu18a}. 

The total interaction potentials V$_T$ (R) = V$_n$ (R) + V$_C$ (R) for the Ni-based reaction 
such as $^{64}$Ni + $^{64}$Ni, $^{64}$Ni + $^{124}$Sn, $^{64}$Ni + $^{132}$Sn, $^{58}$Ni + 
$^{58}$Ni, $^{58}$Ni + $^{124}$Sn, and $^{58}$Ni + $^{132}$Sn systems are obtained for the 
M3Y+EX and R3Y+EX interactions for NL3$^*$ densities. As a representative case, the obtained 
results for total interaction  potentials along with the Coulomb potential V$_C$ and the 
nucleus-nucleus interactions without Coulomb for M3Y+EX and R3Y+EX interactions for 
$^{58}$Ni + $^{58}$Ni system is displayed in Fig. \ref{fig2}. From the figure, we note that 
the nature of the total V$_T$ (R) and the nuclear V$_n$ (R) potentials are similar for both 
the R3Y+EX and M3Y+EX $NN-$interactions (see Fig. \ref{fig2}). Quantitatively, both the 
nuclear potentials obtained from M3Y and R3Y differ significantly particularly in the central 
region and this difference reduced simultaneously with respect to the radial distance. 
Further, the height of the barrier for M3Y $NN-$interaction is a bit higher as compare to 
R3Y case (see the more clearly in the inset of Fig. \ref{fig2}). For example, the R3Y+EX is 
being more attractive by about 1 MeV with a barrier height lower by a few keV, compared to 
the M3Y+EX $NN-$interaction, as is illustrated in the inset of Fig. \ref{fig2}. 

\subsection{Fusion cross-sections}
The barrier characteristics of the nuclear interaction potential i.e. barrier height, 
position and frequency from the total interaction potential are used in Wong formula 
[see Eq. (\ref{eq:27})] for estimating the fusion reaction cross-section for the systems 
such as $^{64}$Ni + $^{64}$Ni, $^{64}$Ni + $^{124}$Sn, $^{64}$Ni + $^{132}$Sn, $^{58}$Ni 
+ $^{58}$Ni, $^{58}$Ni + $^{124}$Sn, and $^{58}$Ni + $^{132}$Sn, known for fusion 
hindrance phenomena. Fig. \ref{fig3} shows the comparison of fusion cross-section obtained 
for $^{64}$Ni + $^{64}$Ni around the Coulomb barrier with the experimental data 
\cite{jian04}. The solid line shows the fusion cross-section using R3Y interaction and 
dashed line using M3Y potential within the Wong formula for NL3$^*$ densities. It is 
observed that R3Y performs relatively superior than M3Y interaction in comparison with 
the experimental data \cite{jian04} below barrier energies. Motivated by this observation, 
the above said calculations are then pursued for $^{64}$Ni + $^{124}$Sn and $^{64}$Ni 
+ $^{132}$Sn reactions as shown in Fig. \ref{fig4}, and Fig. \ref{fig5}, respectively. 
The experimental datas \cite{lesk85,lesk86,jian15,lian07,lian08} are given for comparison. 
For the $^{64}$Ni + $^{124}$Sn reaction, the experimental datas \cite{lesk85,lesk86,jian15} 
are available for near and/or below and other is above Coulomb barrier center of mass 
energies. In Figs. \ref{fig4} and \ref{fig5}, the solid line is for R3Y and dashed for 
M3Y potential. The cross-section corresponding to R3Y are relatively close to the 
experimental data for energies below the Coulomb barrier whereas the M3Y fits the data 
only at above barrier energies. In other words, the nuclear interaction from R3Y potential 
explain the cross-section at comparatively lower energies. It is to be noted that the 
fusion cross-section corresponding to R3Y interaction is always larger as compare to that 
of M3Y potential. Furthermore a few similar calculations are done for another Ni-based 
reactions i.e. $^{58}$Ni + $^{58}$Ni, $^{58}$Ni + $^{124}$Sn and $^{58}$Ni + $^{132}$Sn 
shown in Fig. \ref{fig6}, \ref{fig7}, and \ref{fig8}, respectively with the experimental 
datas \cite{beck81,wolf87,wolf87a,kohl11}. It is clear from all these systems in Figs. 
\ref{fig3}, \ref{fig4}, \ref{fig5}, \ref{fig6}, \ref{fig7}, and \ref{fig8}  that the 
recently developed R3Y interaction within NL3$^*$ force parameter is proven to be 
relatively better choice than M3Y for considering the fusion reactions below the barrier 
at low energies. In other words, R3Y interaction allows the nuclei to relax, which reduces 
the barrier height  and hence increases the fusion cross-section.

The transmission function is accomplished to obtain the fusion barrier distribution 
$\frac{d^2(E \cdot \sigma)}{dE^2}$ by differentiation with respect to center-of-mass energy. 
Classically, the transmission probability is a step function at an energy equal to the 
height of the fusion barrier. The Fermi function blur the step function into a smoother 
function, a parabolic barrier. The $\frac{d^2(E \cdot \sigma)}{dE^2}$ from fusion excitation 
functions are shown in Fig. \ref{fig9} for the R3Y+EX (solid line) and M3Y+EX (dashed 
line) interactions. In the figure, we have shown the fusion barrier distribution for 
reduced cross-sections (a) $^{64}$Ni + $^{64}$Ni, (b) $^{64}$Ni + $^{124}$Sn, 
(c) $^{64}$Ni + $^{132}$Sn, (d) $^{58}$Ni + $^{58}$Ni, (e) $^{58}$Ni + $^{124}$Sn and 
(f) $^{58}$Ni + $^{132}$Sn along with the experimental data 
\cite{jian04,lesk85,lesk86,jian15,lian07,lian08,beck81,wolf87,wolf87a,kohl11} for comparison. 
As expected, here we found the similar predictions as reaction cross-section, the obtained 
results from R3Y are relatively closer to the experimental data for energies below the 
Coulomb barrier whereas the M3Y fits the data only above barrier energies. From the reaction 
cross-section and barrier distribution, one can conclude that the R3Y interaction produce 
relatively better than M3Y potential in comparison to experimental data. Hence, one can 
choice whole microscopic studies using the relativistic mean field density and the recently 
developed relativistic R3Y $NN$ potential for fusion characteristic for above mentioned mass 
region to generate the nuclear potential within double folding procedure.

\section{Summary and Conclusions}
We have investigated possible relationships between the nucleon-nucleon interaction potential 
and the fusion reaction cross-section for a few Ni-based systems, known for fusion hindrance 
phenomena. The fusion barrier distribution for the reduced fusion cross-section are also 
estimated from fusion excitation function for R3Y+EX and M3Y+EX interactions. We have 
considered six reaction systems such as $^{64}$Ni + $^{64}$Ni, $^{64}$Ni + $^{124}$Sn, 
$^{64}$Ni + $^{132}$Sn, $^{58}$Ni + $^{58}$Ni, $^{58}$Ni+$^{124}$Sn and $^{58}$Ni+$^{132}$Sn 
for present analysis. A microscopic approach based on an axial deformed relativistic mean 
field with recently developed NL3$^*$ force has been used along with the Wong formula to 
provide a transparent and analytic way to calculate the fusion cross-section by means of a 
convenient approach to the nucleus-nucleus optical potential. We have considered the well-known 
M3Y and the recently developed relativistic R3Y nucleon-nucleon interaction for estimating the 
nuclear interaction potential. The NL3$^*$ densities for target and projectile are used for 
calculating the nuclear potential within double folding procedure for the study of fusion 
at low energies. It is worth mentioning that the quadrupole, odd multipole (octupole, etc.) 
shape degrees of freedom and/or the corresponding space reflection symmetry may provide some 
of these interesting issues and will throw more light on the fusion properties. We found that 
the R3Y interaction is proven to be better choice than M3Y for considered fusion reactions 
below the barrier energies in prediction of cross-section. Thus it can be inferred that R3Y 
interaction allows interacting nuclei to recline, which leads to lowering the barrier and hence 
increase the cross-section appreciably at the energies below the Coulomb barrier.  The present 
analysis pursue a full microscopic studies by taking the R3Y potential along with the relativistic 
mean field densities within double folding procedure.  

\section*{Acknowledgments}
This work has been supported by FAPESP Project Nos. 2014/26195-5 \& 2017/05660-0, INCT-FNA 
Project No. 464898/2014-5, Seed Money Project of Thapar Institute of Engineering and 
Technology, Department of Science and Technology (DST), Govt. of India Project No. 
YSS/2015/000342 under Young Scientist Scheme, and by the CNPq - Brasil. 


\end{document}